\documentclass[twocolumn,aps,prl,showpacs,superscriptaddress,groupedaddress]{revtex4}
\usepackage[utf8]{inputenc}
\usepackage[T1]{fontenc}
\usepackage{color}
\usepackage{amsmath}
\usepackage{amssymb}
\usepackage{graphicx}
\usepackage{esint}
\usepackage[unicode=true,pdfusetitle,
 bookmarks=true,bookmarksnumbered=false,bookmarksopen=false,
 breaklinks=true,pdfborder={0 0 1},backref=false,colorlinks=true]
 {hyperref}
\usepackage{breakurl}

\makeatletter
\@ifundefined{textcolor}{}
{%
 \definecolor{BLACK}{gray}{0}
 \definecolor{WHITE}{gray}{1}
 \definecolor{RED}{rgb}{1,0,0}
 \definecolor{GREEN}{rgb}{0,1,0}
 \definecolor{BLUE}{rgb}{0,0,1}
 \definecolor{CYAN}{cmyk}{1,0,0,0}
 \definecolor{MAGENTA}{cmyk}{0,1,0,0}
 \definecolor{YELLOW}{cmyk}{0,0,1,0}
}

\usepackage{dcolumn}\usepackage{bm}

\makeatother

\begin{document}

\title{Sculpting a quasi-one-dimensional Bose-Einstein condensate to generate
calibrated matter-waves}

\author{Javed Akram}

\email{javedakram@daad-alumni.de}

\affiliation{Institute f\" ur Theoretische Physik, Freie Universit\" at Berlin, Arnimallee
14, 14195 Berlin, Germany}

\affiliation{Department of Physics, COMSATS, Institute of Information Technology
Islamabad, Pakistan}

\author{Axel Pelster}

\email{axel.pelster@physik.uni-kl.de}

\affiliation{Fachbereich Physik und Forschungszentrum OPTIMAS, Technische Universit\"at
Kaiserslautern, Germany}

\date{\today}
\begin{abstract}
We explore theoretically how to tune the dynamics of a quasi one-dimensional
harmonically trapped Bose-Einstein condensate (BEC) due to an additional
red- and blue-detuned Hermite-Gaussian dimple trap (HGdT). To this
end we study a BEC in a highly non-equilibrium state, which is not
possible in a traditional harmonically confined trap. Our system is
modeled by a time-dependent Gross-Pitaevskii equation, which is numerically
solved by the Crank-Nicolson method in both imaginary and real time.
For equilibrium, we obtain a condensate with two bumps/dips which
are induced by the chosen TEM$_{01}$ mode for the red/blue-detuned
HGdT, respectively. Afterwards, in time-of-flight dynamics, we examine
the adherence/decay of the two bumps/dips in the condensate, which
are induced by the still present red/blue-detuned HGdT, respectively.
On the other hand, once the red/blue HGdT potential is switched off,
shock-waves or bi-trains of gray/dark pair-solitons are created. During
this process it is found that the generation of gray/dark pair-solitons
bi-trains are generic phenomena of collisions of moderately/fully
fragmented BEC. Additionally, it turns out that the special shape
of generated solitons in the harmonically trapped BEC firmly depends
upon the geometry of the HGdT. 
\end{abstract}


\maketitle

\section{Introduction}

The development of laser cooling has accelerated a tremendous interest
in the confinement and manipulation of cold atoms. In particular,
using optical dipole traps generated by red/blue-detuned resonance
laser light has become a versatile tool for the manipulation of atoms.
Recently, optical dipole traps created by a red-detuned laser beam
have become common, experimentally \cite{PhysRevLett.81.2194,PhysRevLett.84.806,PhysRevLett.86.4443,PhysRevLett.87.010404,PhysRevLett.88.020401,PhysRevA.66.051401,PhysRevA.73.043410,PhysRevLett.98.110402,Schulz:2007,PhysRevA.78.033425,Jacob11,PhysRevA.83.013630}
as well as theoretically \cite{PhysRevLett.89.070401,PhysRevA.73.053603,Carpentier08,PhysRevX.1.021003,Uncu11,PhysRevA.84.032322,Javed15},
they are known as a ``tweezer\textquotedblright{} or ``dimple-trap\textquotedblright{}
provided that the trap is quite sharp. Red-detuned dimple traps have
become important tools for BEC production \cite{PhysRevLett.81.2194,PhysRevLett.87.010404,Jacob11},
transport of a BEC over long distances \cite{PhysRevLett.88.020401},
and formation of shock-waves in harmonic plus a dimple trap \cite{Javed15}.
Blue-detuned optical dipole traps, instead, are mostly used as a repulsive
obstacle for atoms \cite{PhysRevLett.83.2502,PhysRevLett.85.2228,PhysRevLett.87.210402,PhysRevLett.104.160401,Javed15}.
The HG laser beams are higher order solutions of the paraxial wave
equation with rectangular symmetry about their axes of propagation
\cite{saleh2007fundamentals,milonni2010laser}. Due to their enormous
application areas, there have been several attempts to develop such
higher order beam modes \cite{Kozawa:05,Flores-Perez:06,Novitsky06}.
For example, the resonator of a laser is manipulated such that the
beam is emitted in a desired beam mode structure \cite{Mushiake72},
or transforms a general Gaussian laser-beam with interferometric methods
into the desired modes \cite{PhysRevLett.91.233901,Torok:04,Toussaint:05}.
These interferometric methods are typically based on the addition
or subtraction of different scalar laser beam modes \cite{Tidwell:93}.
To switch between different modes, a more flexible way is to use a
spatial light modulator to generate the desired higher order laser
modes \cite{Stalder:96,Neil:02}. It is already known that HG laser
modes possess interesting properties, however, to the best of our
knowledge, so far only two experimental papers have been published
about the confinement of atoms in higher order optical dipole traps
\cite{Meyrath:05,Smith05}.

In this paper, we consider a theoretical analysis of a quasi one-dimensional
(1D) Bose-Einstein condensate confined by both a harmonic trap and
a Hermite-Gaussian dimple trap (HGdT). The red/blue-detuned HGdT can
be generated by using the Hermite-Gaussian (HG) laser beam. The mean-field
description of the one-dimensional macroscopic BEC wave function is
based upon the Gross-Pitaevskii equation \cite{Pitaevskii03,Pethick02,Kevrekidis08}.
A truly 1D mean-field regime, also known as Tonks-Girardeau regime,
requires transverse dimensions of the trap on the order of or less
than the atomic s-wave scattering length \cite{PhysRevLett.81.938,PhysRevLett.85.3745,PhysRevLett.91.163201}.
In contrast, the quasi one-dimensional regime of the Gross-Pitaevskii
equation holds when the transverse dimension of the trap is larger
than or of the order of the s-wave scattering length and much smaller
than the longitudinal extension \cite{PhysRevA.57.3837,PhysRevA.58.2417,Lincoln00,PhysRevA.62.063611,Adhikari06}.
Here we focus our attention to a quasi one-dimensional Gross-Pitaevskii
equation (1DGPE). This regime is quite interesting, as it is well-known
to feature bright solitons for attractive s-wave scattering lengths
\cite{Strecker03,Abdullaev04,Herring2005144,Becker08}, gray/dark
solitons for repulsive s-wave scattering lengths \cite{Denschlag07012000,PhysRevA.65.053605,PhysRevA.70.013602,Yan12,PhysRevA.92.013627}
or the formation of shock waves in a BEC \cite{PhysRevLett.101.170404,PhysRevA.80.043606}.

With this, we organize our paper as follows. We derive the underlying
quasi one-dimensional Gross-Pitaevskii equation (1DGPE) in Sec.~\ref{Sec2},
where, we also outline the system geometry and relate our simulation
parameters to tunable experimental parameters. Afterwards in Sec.~\ref{Sec3},
for the equilibrium properties of the system, we compare a Thomas-Fermi
approximate solution with numerical results and show that the HGdT
imprint upon the condensate wave function strongly depends upon whether
the HGdT is red or blue-detuned. Later, in Sec.~\ref{Sec4} we assume that
the magnetic trap is switched off and we determine the time-of-flight
(TOF) dynamics of the condensate wave function, when the HGdT is still
present. On the one hand we obtain that for red-detuning the HGdT
imprint does not decay, but for the blue-detuning the HGdT imprint
decreases during TOF. On the other hand, we discuss in detail how
the collision of the condensate with the HGdT potential during the
non-ballistic expansion leads to characteristic matter-wave stripes.
In Sec.~\ref{Sec5}, we investigate instead matter-wave interferences in form
of the formation of shock-waves/gray(dark) pair-soliton bi-trains
in the harmonic trap, after having switched off the red/blue-detuned
HGdT potential. There, we also find out that the generation of gray/dark
pair-solitons bi-trains represents a generic phenomenon of collisions
of moderately/fully fragmented BEC, which strongly depends upon the
equilibrium values of the red/blue-detuned HGdT depth. Finally, Sec.~\ref{Sec6}
provides a summary and conclusions.

\section{modified quasi 1D model}
 
\begin{figure*}
\includegraphics[width=18cm,height=11cm]{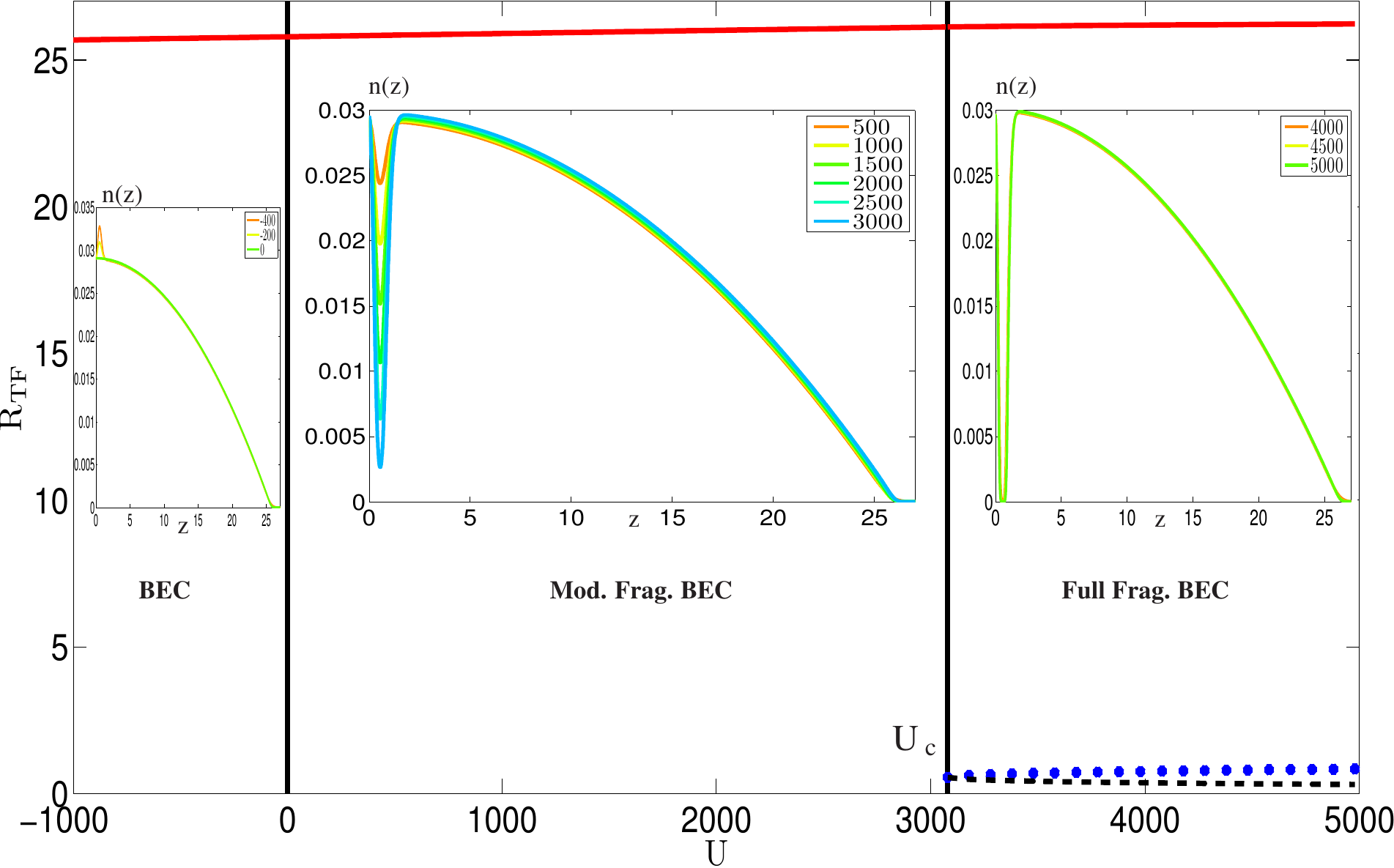} \protect\protect\protect\protect\caption{(Color online) Outer Thomas-Fermi radius $\text{R}_{\text{TF1}}$
(red solid), central Thomas-Fermi radius $\text{R}_{\text{TF2}}$
(blue dashed) and inner Thomas-Fermi radius $\text{R}_{\text{TF3}}$
(dotted black) as a function of HGdT depth $U$ in dimensionless units
for the coupling constant value $G_{\text{B}}=11435.9$. Below
the critical value HGdT depth $U<U_{\text{c}}$, the
BEC is moderately (Mod.) fragmented (Frag.), with one exceptional
case at $U=0$, where the BEC is completely confined in one-dimensional
harmonic trap. The BEC fully fragments into three parts above the
critical value $U_{\text{c}}\approx3079$, as can be seen in
the specific regional inset density plots. \label{Fig1}}
\end{figure*}

\label{Sec2}  We consider a one-component BEC with time-dependent two-particle interactions
described by the three-dimensional GPE 
\begin{align}
i\hbar\frac{\partial}{\partial t}\psi\left(\mathbf{r},t\right)=\left\{ -\frac{\hbar^{2}}{2m_{B}}\nabla^{2}+V(\mathbf{r})+U_{\text{dT}}^{\text{3D}}\right.\label{eq1}\\
\left.+G_{\text{B}}^{\text{3D}}\parallel\psi\left(\mathbf{r},t\right)\parallel^{2}\right\} \psi\left(\mathbf{r},t\right),\nonumber 
\end{align}
where $\psi\left(\mathbf{r},t\right)$ denotes the macroscopic condensate
wave function for the $^{87}\text{Rb}$ BEC with the spatial coordinates
$\mathbf{r}=\left(x,\; y,\; z\right)$. Here $m_{\text{B}}$ stands
for the mass of the $^{87}\text{Rb}$ atom, $G_{\text{B}}^{\text{3D}}=N_{\text{B}}4\pi\hbar^{2}a_{\text{B}}/m_{\text{B}}$
represents the three-dimensional $^{87}\text{Rb}$ coupling constant,
where $N_{\text{B}}=20\times10^{4}$ denotes the number of $^{87}\text{Rb}$
atoms, and the s-wave scattering length is $a_{\text{B}}=94.7~{\rm {a}_{0}}$
with the Bohr radius ${\rm {a}_{0}}$. Furthermore, $V\left(\mathbf{r}\right)=m_{\text{B}}\omega_{\text{z}}^{2}z^{2}/2+m_{\text{B}}\omega_{\text{r}}^{2}\left(x^{2}+y^{2}\right)/2$
describes a three-dimensional harmonic confinement, which has rotational
symmetry with respect to the $z$-axis. The oscillator lengths for
experimental parameters are $l_{\text{z}}=\sqrt{\hbar/m_{\text{B}}\omega_{\text{z}}}=4.12\,\mu\text{m}$
and $l_{\text{r}}=\sqrt{\hbar/m_{\text{B}}\omega_{\text{r}}}=0.84\,\mu\text{m}$
for the trap frequencies $\omega_{\text{z}}=2\pi\times6.8\,\text{Hz}$
and $\omega_{\text{r}}=2\pi\times160\,\text{Hz}$, respectively.

An additional three-dimensional narrow Hermite-Gaussian laser beam
polarizes the neutral atoms which yields the HGdT potential $U_{\text{dT}}^{\text{3D}}=U_{\text{0}}\text{I}_{nm}\left(\mathbf{r}\right)$.
Within the rotating-wave approximation its amplitude is $U_{\text{0}}=3\pi c^{2}\Gamma/\left(2\omega_{\text{A}}^{3}\Delta\right)$
\cite{Allen1987,ScullyZubairy,saleh2007fundamentals,milonni2010laser},
where $\Gamma=\left|<e|{\bf d}|g>\right|^{2}\omega_{\text{A}}^{3}/\left(3\pi\epsilon_{0}\hbar c^{3}\right)$
denotes the damping rate due to energy loss via radiation, which is
detected by the dipole matrix element between ground and excited state.
Furthermore, $\Delta=\omega-\omega_{\text{A}}$ represents the detuning
of the laser, here $\omega$ is the laser frequency and $\omega_{\text{A}}$
stands for the atomic frequency. And $\text{I}_{nm}\left(\mathbf{r}\right)$
describes the intensity profile of the TEM$_{nm}$ Hermite-Gaussian
laser beam, which is assumed to propagate in $y$-direction and is
determined via 
\begin{eqnarray}
\text{I}_{nm}\left(\mathbf{r}\right)= & \frac{2P}{2^{n+m}n!m!\pi}H_{n}\left(\frac{\sqrt{2}x}{W_{\text{x}}\left(y\right)}\right)^{2}H_{m}\left(\frac{\sqrt{2}z}{W_{\text{z}}\left(y\right)}\right)^{2}\label{eq2}\\
 & \times\frac{e^{-\left[\frac{2x^{2}}{W_{\text{x}}^{2}\left(y\right)}+\frac{2z^{2}}{W_{\text{z}}^{2}\left(y\right)}\right]}}{W_{\text{x}}\left(y\right)W_{\text{z}}\left(y\right)}\,,\nonumber 
\end{eqnarray}
with $P=\int\int\text{I}_{nm}\left(\mathbf{r}\right)dxdz$ being the
normalization constant. Furthermore $W_{\text{x/z}}^{2}(y)=W_{\text{0x/z}}^{2}\left(1+y^{2}/y_{\text{Rx/z}}^{2}\right)$
denotes the Gaussian beam radius in the $x$- and $z$-direction,
where the intensity decreases to $1/e^{2}$ of its peak value, $y_{Rx/z}=\pi W_{0x/z}^{2}/\lambda$
represents the so-called Rayleigh-lengths, which are defined as the
distance from the focus $W_{\text{0x/z}}$ position where the beam
radius increases by a factor of $\sqrt{2}$ \cite{saleh2007fundamentals}.
Here $H_{n}\left(q\right)$ and $H_{m}\left(q\right)$ are Hermite
polynomials of order $n$ and $m$ in $x$- and $z$-directions, respectively.
In the following we restrict ourselves to a HGdT potential for a BEC,
which is based on a Hermite-Gaussian TEM$_{01}$ laser beam mode and
thus carries a dark spot in the center of the profile: 
\begin{align}
\text{I}_{01}\left(\mathbf{r}\right)=\frac{8Pz^{2}}{\pi W_{\text{x}}\left(y\right)W_{\text{z}}^{3}\left(y\right)}e^{-\left[\frac{2x^{2}}{W_{\text{x}}^{2}\left(y\right)}+\frac{2z^{2}}{W_{\text{z}}^{2}\left(y\right)}\right]}\,.\label{eq3}
\end{align}
For the TEM$_{01}$ laser beam, we use the width along the $x$-axis
$W_{\text{0x}}=1.1\,\mu\text{m}$ and along the $z$-axis $W_{\text{0z}}=3.2\,\mu\text{m}$.
Therefore, the Rayleigh lengths for the red-detuned laser light with
$\lambda=840\,\text{nm}$ \cite{PhysRevA.83.013630} yield $y_{\text{Rx}}=4.526\,\mu\text{m}$
and $y_{\text{Rz}}=38.29\,\mu\text{m}$ as well as for the blue detuned
laser light with $\lambda=772\,\text{nm}$ \cite{Xu:10} we get $y_{\text{Rx}}=4.92\,\mu\text{m}$
and $y_{\text{Rz}}=41.6\,\mu\text{m}$. With keeping in mind the fact
$y_{\text{Rx}/\text{z}}\gg l_{\text{r}}$, we can approximate the
widths of the HG laser beam in $x$- and $z$-direction according
to $W_{\text{x}/\text{z}}(y)\approx W_{0\text{x}/\text{z}}$. This
simplifies the HGdT potential to 
\begin{align}
U_{\text{dT}}^{\text{3D}}\left(\mathbf{r}\right) & =\frac{8 U_{0}Pz^{2}}{\pi W_{\text{0x}}W_{\text{0z}}^{3}}e^{-\left(\frac{2x^{2}}{W_{0\text{x}}^{2}}+\frac{2z^{2}}{W_{0\text{z}}^{2}}\right)}.\label{eq4}
\end{align}
As we have an effective one-dimensional setting due to $\omega_{\text{z}}\ll\omega_{\text{r}}$,
which implies $l_{\text{z}}>l_{\text{r}}$, and $y_{\text{Rx}/\text{z}}\gg l_{\text{r}}$,
we factorize the BEC wave-function via $\psi(\mathbf{r},t)=\psi(z,t)\phi(r_{\perp},t)$
with ${\bf r}_{\perp}=\left(x,\; y\right)$ and %
\begin{eqnarray}
\phi({\bf r}_{\perp},t) & = & \frac{e^{-\frac{x^{2}+y^{2}}{2l_{\text{r}}^{2}}}}{\sqrt{\pi}l_{\text{r}}}e^{-i\omega_{\text{r}}t}\,.\label{eq5}
\end{eqnarray}
We follow Ref.~\cite{Kamchatnov04} and integrate out the two transversal
dimensions of the three-dimensional GPE. After some algebra, the resulting
quasi one-dimensional GPE reads 
\begin{align}
i\hbar\frac{\partial}{\partial t}\psi\left(z,t\right)=\left\{ -\frac{\hbar^{2}}{2m_{B}}\frac{\partial^{2}}{\partial z^{2}}+V\left(z\right)+Uz^{2}e^{-\frac{2z^{2}}{W_{\text{0z}}^{2}}}\right.\label{eq6}\\
\left.+G_{\text{B}}\parallel\psi\left(z,t\right)\parallel^{2}\right\} \psi\left(z,t\right),\nonumber 
\end{align}
where $V\left(z\right)=m_{\text{B}}\omega_{z}^{2}z^{2}/2$ represents
an effective one-dimensional harmonic potential from the MOT, and
the one-dimensional two-particle interaction strength turns out to
be 
\begin{equation}
G_{\text{B}}=2N_{\text{B}}a_{\text{B}}\hbar\omega_{\text{r}}\,.\label{eq7}
\end{equation}
Furthermore, the one-dimensional HGdT depth results in 
\begin{equation}
U=\frac{8U_{0}P}{\pi W_{\text{0z}}^{3}\sqrt{W_{\text{0x}}^{2}+2l_{\text{r}}^{2}}}\,,\label{eq8}
\end{equation}

In order to make the 1DGPE in (\ref{eq6}) dimensionless, we introduce
the dimensionless time as $\tilde{t}=\omega_{\text{z}}t$, the dimensionless
coordinate $\tilde{z}=z/l_{\text{z}}$, and the dimensionless wave
function $\tilde{\psi}=\psi\sqrt{l_{\text{z}}}$. With this Eq.~(\ref{eq6})
can be written in dimensionless form 
\begin{align}
i\frac{\partial}{\partial\tilde{t}}\tilde{\psi}\left(\tilde{z},\tilde{t}\right)=\left\{ -\frac{1}{2}\frac{\partial^{2}}{\partial\tilde{z}^{2}}+\frac{1}{2}\tilde{z}^{2}+\tilde{U}\tilde{z}^{2}e^{-\frac{\tilde{z}^{2}}{\tilde{\alpha}^{2}}}\right.\label{eq9}\\
\left.+\tilde{G}_{\text{B}}\parallel\tilde{\psi}\left(\tilde{z},\tilde{t}\right)\parallel^{2}\right\} \tilde{\psi}\left(\tilde{z},\tilde{t}\right),\nonumber 
\end{align}
here $\tilde{G}_{\text{B}}=2N_{\text{B}}\omega_{\text{r}}a_{\text{B}}/\omega_{\text{z}}l_{\text{z}}$,
and $\tilde{U}=8U_{0}Pl_{\text{z}}/\left(\pi\omega_{\text{z}}W_{\text{0z}}^{3}\sqrt{W_{\text{0x}}^{2}+2l_{\text{r}}^{2}}\right)$
are the dimensionless two-particle coupling strength and the dimensionless
HGdT depth, respectively. The above mentioned experimental values
yield the dimensionless Rb-Rb coupling constant $\tilde{G}_{\text{B}}=11435.9$
and $\tilde{\alpha}=W_{\text{0z}}/\left(\sqrt{2}l_{\text{z}}\right)=0.548$
represents the ratio of the width of the HGdT potential and the harmonic
oscillator length along the $z$-axis. Furthermore, the typical depth
of dipole potential traps ranges from micro-kelvin to nano-kelvin
\cite{Bongs04,Yin20061}, which yields $\tilde{U}$ to be of
the order of up to few thousands. From here on, we will drop all tildes
for simplicity.

\begin{figure*}
\includegraphics[scale=0.8]{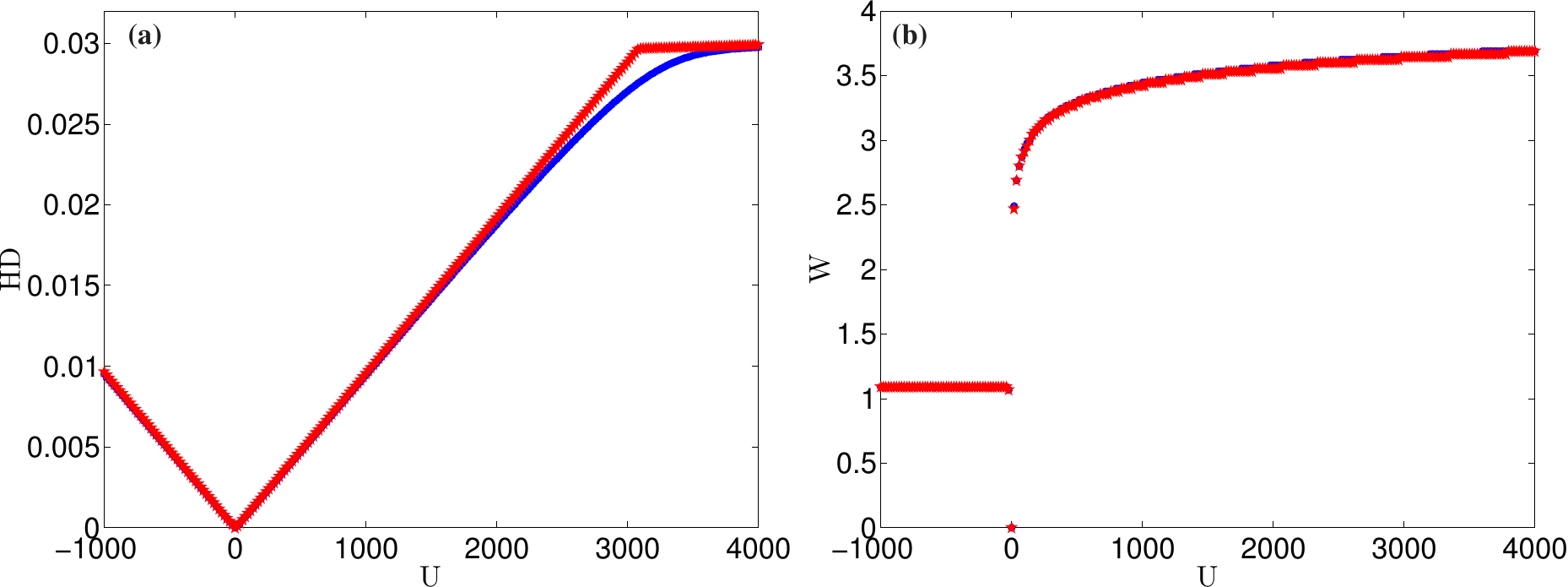} \protect\protect\protect\protect\caption{(Color online) (a) Height/depth and (b) width of red/blue-detuned HGdT
induced bumps/dips according to Eq.~(13) and $\text{W}=2z_{\text{Max}}$,
respectively, versus red/blue-detuned HGdT depth $U$ in dimensionless
units for the experimental BEC coupling constant $G_{\text{B}}=11435.9$
determined numerically by solving the quasi 1DGPE (\ref{eq9}) in
dimensionless imaginary time (blue circles) and analytically (red
stars) from the BEC TF wave function (\ref{eq10}). \label{Fig2} }
\end{figure*}

\section{Stationary Condensate Wave Function}
\label{Sec3}
In order to determine the equilibrium properties of the red/blue-detuned
HGdT potential imprint on the condensate wave function, we solve the
quasi 1DGPE~(\ref{eq9}) numerically by using the split-operator
method in imaginary time \cite{Vudragovic12,Kumar15,Loncar15,Sataric16}.
The HGdT imprint induces two bumps/dips at the center of the BEC density
for red/blue-detuned HGdT as shown in the insets of Fig.~\ref{Fig1}.
For stronger red-detuned HGdT depth values the two bumps increase
further, but for stronger blue-detuned HGdT depth the two dips in
the BEC density get deeper and deeper until the BEC fragments into
three parts as shown in the inset of Fig.~\ref{Fig1}. To investigate
this scenario in more detail, we argue that, due to $G_{\text{B}}\gg1$,
the TF approximation is valid, as the inequality ${\rm E_{\text{{\rm int,pot}}}/E_{\text{{\rm kin}}}\gg1}$
holds within the whole region of interest for the HGdT depth $U$ \cite{Javed15}.

Therefore we perform for the condensate wave function the ansatz $\psi(z,t)=\psi(z)e^{-i\mu t}$,
insert it into the modified quasi 1DGPE (\ref{eq9}), and neglect
the kinetic energy term, yielding the density profile 
\begin{eqnarray}
\psi\left(z\right)=\sqrt{\frac{\mu}{G_{\text{B}}}\left(1-\frac{z^{2}}{2\mu}-\frac{Uz^{2}}{\mu}e^{-\frac{z^{2}}{\alpha^{2}}}\right)}\label{eq10}\\
\times\Theta\left(1-\frac{z^{2}}{2\mu}-\frac{Uz^{2}}{\mu}e^{-\frac{z^{2}}{\alpha^{2}}}\right)\,.\nonumber 
\end{eqnarray}
In view of the normalization $2\intop_{0}^{\infty}\parallel\psi\left(z\right)\parallel^{2}dz=1$,
which fixes the chemical potential $\mu$, we determine the Thomas-Fermi
radii $\text{R}_{\text{TF}}$ from the condition that the condensate
wave function vanishes:

\begin{figure*}
\includegraphics[scale=0.9]{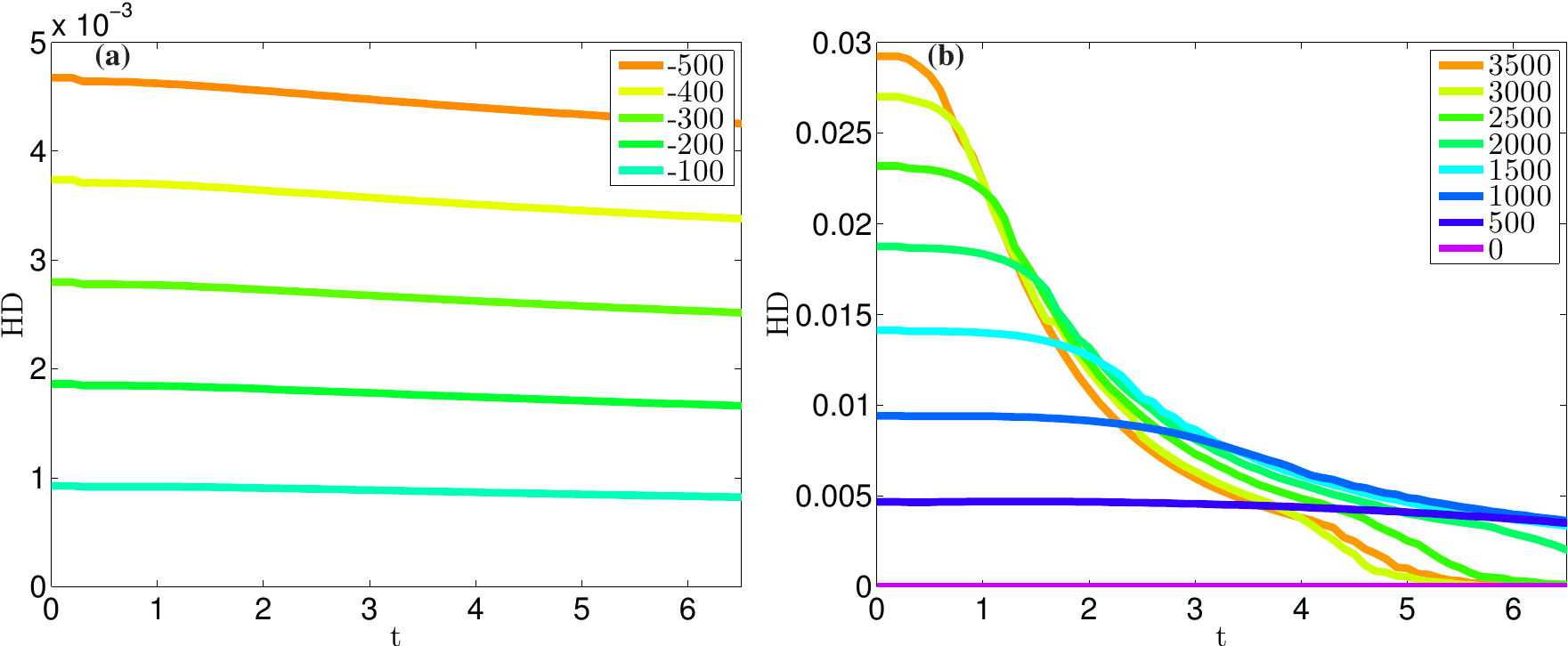} \protect\protect\protect\protect\caption{(Color online) Height/depth versus dimensionless time during TOF for
different (a) red-detuned and (b) blue-detuned HGdT depths $U$ in dimensionless
units.\label{Fig3} }
\end{figure*}

\begin{figure}
\includegraphics[width=9cm,height=20cm]{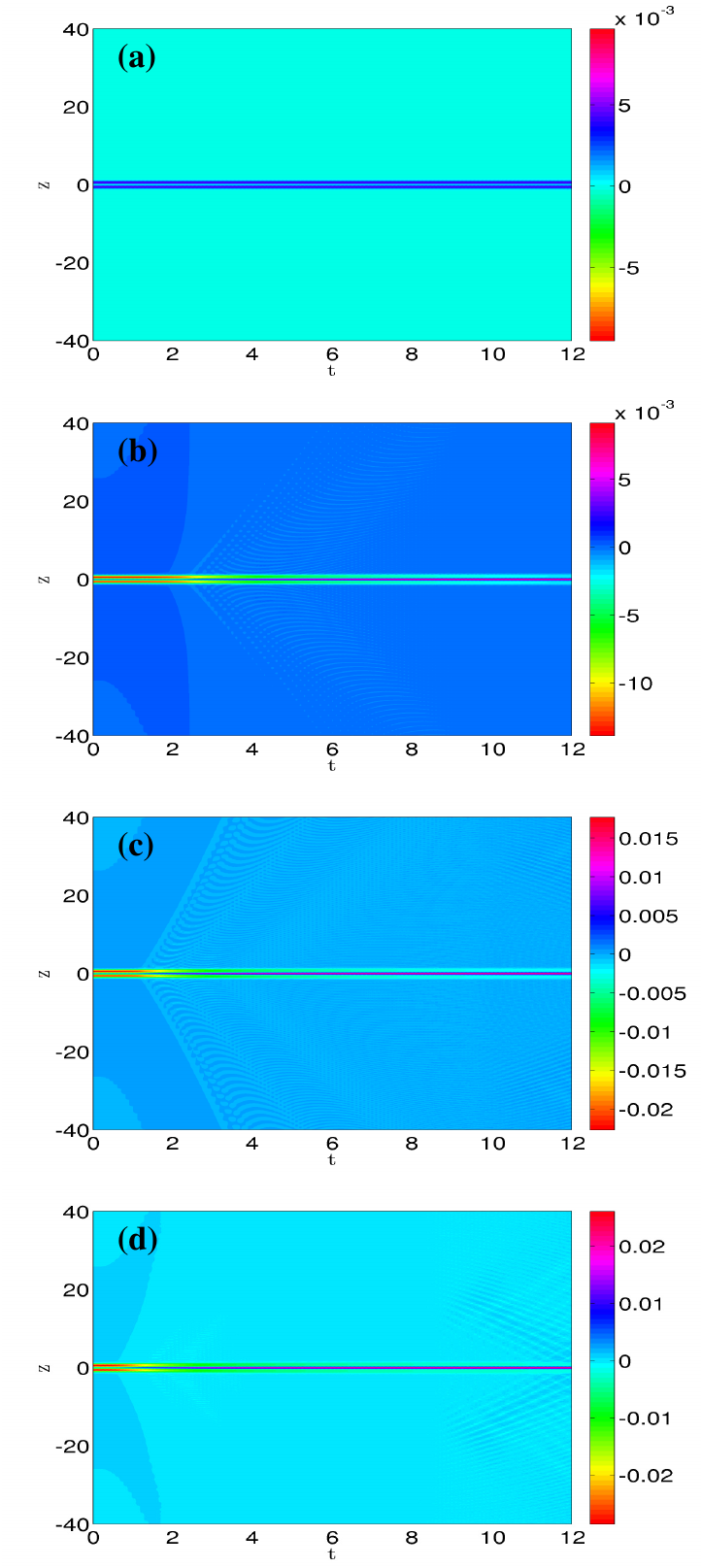} \protect\protect\protect\protect\caption{(Color online) Time-of-flight evolution of depleted density $\parallel\psi\left(z,t\right)\parallel_{\text{DD}}^{2}=\parallel\psi\left(z,t\right)\parallel_{U\protect\neq0}^{2}-\break\parallel\psi\left(z,t\right)\parallel_{U=0}^{2}$
from solving the modified quasi 1DGPE equation for different values
of $|U|$: (a) $U=-500$, (b) $U=1500$, (c) $U=2500$,
and (d) $U=3500$ in dimensionless units.\textbf{ }\label{Fig4} }
\end{figure}

\begin{eqnarray}
\mu=\frac{\text{R}_{\text{TF}}^{2}}{2}+U\text{R}_{\text{TF}}^{2}e^{-\frac{\text{R}_{\text{TF}}^{2}}{\alpha^{2}}}\,.\label{eq11}
\end{eqnarray}
As can be read off from the inset of the Fig.~\ref{Fig1} the number
of solutions of Eq.~(\ref{eq10}) changes for increasing red/blue-detuned
HGdT depth U. In the case, when $U$ is smaller than $U_{\text{c}}$,
Eq.~(\ref{eq10}) defines only the BEC cloud radius $\text{R}_{\text{TF1}}$.
But for the case $U>U_{\text{c}}$, the blue-detuned
HGdT drills two holes at the center of the $^{87}\text{Rb}$ condensate,
so the BEC fragments into three parts as shown in Fig.~\ref{Fig1}.
Thus, we have then, apart from the outer condensate radius $\text{R}_{\text{TF1}}$,
also two inner condensate radii $\text{R}_{\text{TF2}}$ and $\text{R}_{\text{TF3}}$.
With this the normalization condition $2\intop_{0}^{\text{R}_{\text{TF3}}}\parallel\psi\left(z\right)\parallel^{2}dz+2\intop_{\text{R}_{\text{TF2}}}^{\text{R}_{\text{TF1}}}\parallel\psi\left(z\right)\parallel^{2}dz=1$
yields 
\begin{eqnarray}
\mu\left(\text{R}_{\text{TF1}}-\text{R}_{\text{TF2}}+\text{R}_{\text{TF3}}\right)-\frac{1}{6}\left(\text{R}_{\text{TF1}}^{3}-\text{R}_{\text{TF2}}^{3}+\text{R}_{\text{TF3}}^{3}\right)\,\,\,\label{eq12}\\
+\frac{\alpha^{2}U}{2}\left(\text{R}_{\text{TF1}}e^{-\frac{\text{R}_{\text{TF1}}^{2}}{\alpha^{2}}}-\text{R}_{\text{TF2}}e^{-\frac{\text{R}_{\text{TF2}}^{2}}{\alpha^{2}}}+\text{R}_{\text{TF3}}e^{-\frac{\text{R}_{\text{TF3}}^{2}}{\alpha^{2}}}\right)=\nonumber \\
\frac{\alpha^{3}\sqrt{\pi}U}{4}\left[\text{Erf}\left(\frac{\text{R}_{\text{TF1}}}{\alpha}\right)-\text{Erf}\left(\frac{\text{R}_{\text{TF2}}}{\alpha}\right)+\text{Erf}\left(\frac{\text{R}_{\text{TF3}}}{\alpha}\right)\right]\,,\nonumber 
\end{eqnarray}
where $\text{Erf}(y)=\frac{2}{\sqrt{\pi}}\int_{0}^{y}e^{-x^{2}}dx$
denotes the error function. In case of $U<U_{\text{c}}$,
the BEC's inner two radii $\text{R}_{\text{TF2}}$ and $\text{R}_{\text{TF3}}$
vanish and the BEC outer radius is approximated via $\text{R}_{\text{TF1}}\approx\sqrt{2\mu}$
due to Eq.~(\ref{eq11}). Thus, for $U<U_{\text{c}}$
the BEC chemical potential is determined explicitly from (\ref{eq12}):
$\mu\approx3^{2/3}/2^{7/2}\left(2G_{\text{B}}+\sqrt{\pi}\alpha^{3}U\right)^{2/3}$.
Provided that $U>U_{\text{c}}$, two inner cloud radii
$\text{R}_{\text{TF2}}$ and $\text{R}_{\text{TF3}}$ have to be taken
into account according to Fig.~\ref{Fig1}. We observe that the Thomas-Fermi
value of the critical red/blue-detuned HGdT depth $U_{\text{c}}\approx3079$
is close to the numerical one $U_{\text{c}}\approx3090$. Figure~\ref{Fig1}
also shows the resulting outer and inner Thomas-Fermi radius as a
function of the red/blue-detuned HGdT depth U. Here, the two inner
radii behave symmetric, e.g., for $U>3079$ the $\text{R}_{\text{TF2}}$
is increasing and $\text{R}_{\text{TF3}}$ is decreasing correspondingly,
however after $U\gtrsim4500$, they both become approximately
constant as shown in Fig.~\ref{Fig1}. We also read off that $\text{R}_{\text{TF1}}\approx\sqrt{2\mu}$
remains approximately constant for $U>U_{\text{c}}$,
so we conclude that the chemical potential $\mu$ is then sealed to
its critical value $\mu_{c}\approx341.28$.

In the perspective of a quantitative comparison between the analytical
and the numerical calculation, we characterize the red/blue-detuned
HGdT induced imprint upon the condensate wave function by the following
two quantities. The first one is the red/blue-detuned HGdT induced
imprint height/depth

\begin{eqnarray}
 & \begin{array}{r@{}l@{\,}l}
\text{HD}=\left\{ \begin{array}{r@{}l@{\,}l}
\text{Max}\left(\parallel\Psi\left(z\right)\parallel^{2}\right)-\parallel\Psi\left(0\right)\parallel^{2}\quad U\leq0\vspace{5mm}\label{eq13}\\
\parallel\Psi\left(0\right)\parallel^{2}-\text{Min}\left(\parallel\Psi\left(z\right)\parallel_{z\rightarrow0}^{2}\right)\quad U>0
\end{array}\right.\end{array}
\end{eqnarray}
and the second one is the red/blue-detuned HGdT induced imprint width
$\text{W}=2z_{\text{Max}},$ where $z_{\text{Max}}$ denotes the coordinate
of maximal density. To find out a one-to-one resemblance between analytical
and numerical calculation of HD and W, we determine the solution of
the dimensionless 1DGPE~(\ref{eq9}) and compare it with the TF solution
of Eq.~(\ref{eq10}), as shown in Fig.~\ref{Fig2}. The case $U=0$,
i.e., when the HGdT potential is switched off, corresponds to a BEC
in a quasi one-dimensional harmonic trap. Furthermore, in the range
$U<U_{\text{c}}$ we observe that the red/blue-detuned
HGdT induced imprint height/depth changes linearly with the optical
dipole trap depth $U$ according to 
\begin{eqnarray}
\text{HD}\approx\frac{\alpha^{2}|U|e^{T\left(-e/2|U|\right)-1}}{G_{\text{B}}}\, & U\neq0\,,\alpha\neq0\,,\label{eq14}
\end{eqnarray}
where $T=xe^{x}$ abbreviates the productlog function. In case of
$U>U_{\text{c}}$ the blue-detuned HGdT induced imprint
depth yields the constant value $\text{\text{HD}}_{\text{c}}\approx0.0296$
as follows from Eq.~(\ref{eq14}), which slightly deviates from the
corresponding numerical value $\text{\text{HD}}_{\text{c}}=0.027$.
Similarly, the red/blue-detuned HGdT induced imprint width follows
from $\text{W}=2\alpha\sqrt{1-T\left(-e/2|U|\right)}$ according
to the TF approximation, which reduces at the critical blue-detuned
optical dipole depth to $\text{W}_{\text{c}}\approx3.60$ whereas
the corresponding numerical value is $\text{W}_{\text{c}}\approx3.64$,
as shown in Fig.~\ref{Fig2}.

\section{Time-of-Flight dynamics of red/blue-detuned HGdT induced imprint}
\label{Sec4}
The time-of-flight (TOF) expansion has been used to measure various
BEC properties since the field's inception. By suddenly turning off
the magnetic trap, when the HGdT is still present, the atom cloud
is allowed to expand in all directions. This expansion proceeds according
to the momenta of the atoms at the initial time $t=0$ and an additional
tiny acceleration results from inter-particle interactions. The red-detuned
HGdT induced two bumps remain approximately constant during the temporal
evolution as shown in Fig.~\ref{Fig3}(a). But the blue-detuned HGdT
induced two dips at the center of the condensate start decaying with
a characteristic time scale after having switched off the trap as
shown in Fig.~\ref{Fig3}(b). Furthermore, the dips of the HGdT induced
imprint start decaying faster with increasing blue-detuned HGdT depth
for smaller time as shown in Fig.~\ref{Fig3}(b). Note that the relative
speed of the bumps or dips from each other turns out to vanish.

Furthermore, we investigate in detail the possible occurrence of matter-wave
stripes at the top of the condensate during the non-ballistic expansion
of the moderately/fully fragmented BEC cloud by plotting the density
distribution of the released cloud correspondingly as shown in Fig.~\ref{Fig4}.
According to Fig.~\ref{Fig4}(a) we do not observe any particular
structure for the red-detuned HGdT induced imprint, but Fig.~\ref{Fig4}(b-d)
shows for the blue-detuned HGdT induced imprint that characteristic
matter-wave stripes occur, which are generated, while the freely expanding
BEC collides with the HGdT potential. For small blue-detuned HGdT
depth, the generation of matter-wave stripes can be seen at later
time, as compared to higher blue-detuned HGdT depth, as shown in Fig.~\ref{Fig4}(b-d).
The matter-wave stripes are directly visible for $U<U_{\text{c}}$,
as can be explained as follows. In Fig.~\ref{Fig4}(b), the height
of the blue-detuned HGdT induced two dips is smaller as compared to
Fig.~\ref{Fig4}(c), therefore they need more time to drill a hole
in the condensate during TOF. For the HGdT potential depth $U=1500$,
the BEC fragments into three parts at the dimensionless time $t=2.4$,
afterwards the three fragmented condensates start to interact as separate
identities with the HGdT potential, which leads to the formation of
characteristic matter-wave stripes. The similar phenomenon happens
in Fig.~\ref{Fig4}(c), but in this example the initial HGdT potential
depth $U=2500$ is larger than the previous one in Fig.~\ref{Fig4}(b),
so the BEC becomes fragmented at the earlier time $t=1.3$. In the example
of Fig.~\ref{Fig4}(d), when $U>U_{\text{c}}$, the
BEC is already initially, i.e. at time $t=0$, fragmented according
to Fig.~\ref{Fig1}. Therefore the matter-wave stripes can be seen
just after $t>0$, but the stripes are not as visible as in the two
previous cases.

\begin{figure*}
\includegraphics[width=18cm,height=10cm]{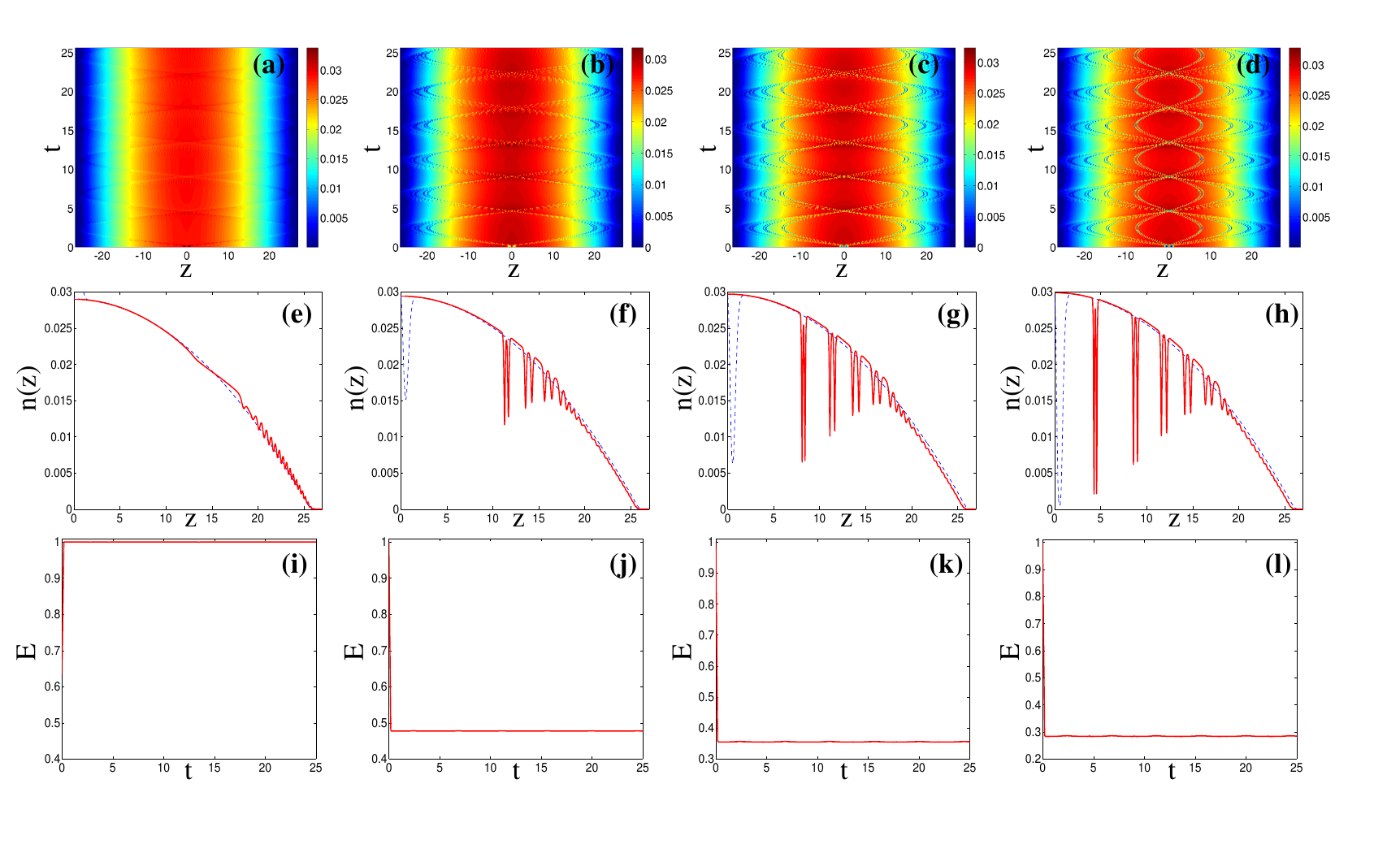}
\caption{(Color online) Coherent matter-waves evolution in the BEC density,
when red/blue-detuned HGdT potential is switched off, versus time
and position for different values of (a) $U=-500$, (b) $U=1500$,
(c) $U=2500$, and (d) $U=3500$. Graphs (e-h) show the
BEC density at time $t=5.3$, and in graphs (i-l) we plotted the corresponding
energies $E(t)= E(t)/\text{Max}( E(t))$ versus
time for the HGdT depths $U$ of the graphs (a-d) in dimensionless
units. \label{Fig5} }
\end{figure*}

\section{Shock-waves and gray/dark pair-solitons bi-trains}
\label{Sec5}
In this section, we show that matter-wave self-interferences emerge,
once the red/blue-detuned HGdT potential is suddenly switched off,
within the remaining harmonic confinement, as this leads to shock
waves and gray/dark pair-solitons bi-trains, respectively as shown
in Fig.~\ref{Fig5}(a-d). A shock-wave is a special kind of propagating
disturbance in the BEC, whose amplitude, unlike for solitons, decreases
relatively quickly with large distance. Furthermore, gray/dark solitons
have a characteristic property that they can pass through one another
without any change of shape, amplitude, or speed. We can see from
Fig.~\ref{Fig5}(b-d) that the pair-solitons bi-trains do, indeed,
pass through one another and that they are reflected from the end
of the trapping potential.

Once the red/blue-detuned HGdT potential is switched off, the system
quasi-instantaneously adjusts its energy to the new equilibrium, paving
the way for the creation of shock-waves and bi-trains of gray/dark
pair-solitons, respectively. The total normalized energy ${\rm E({\rm t)={\rm E(t)/\text{Max}({\rm E(t))}}}}$
as shown in Fig.~\ref{Fig5}(i-l), changes quite quickly from its
initial value to a new equilibrium value, thus generating the shock-waves
or the pair-solitons bi-trains. For an initial red- and blue-detuned
HGdT depth, we observe that two excitations of the condensate are
created at the position of the red/blue-detuned HGdT potential, which
travel in the opposite direction with the same center-of-mass speed,
are reflected back from the harmonic trap boundaries, and then collide
at the red/blue-detuned HGdT potential position as shown in Fig.~\ref{Fig5}(a-d).

We have performed calculations for different red-detuned HGdT potential
depths $U<0$ and in all cases we observe the formation of
the shock-wave structures as shown in Fig.~\ref{Fig5}(a). The density
of atoms around the shocks is mostly enhanced in comparison with the
density far away from these perturbations. And for the blue-detuned
HGdT potential trap, we detect gray/dark pair-solitons bi-trains,
traveling in opposite directions with the same speed as shown in Fig.~\ref{Fig5}(b-d).
The creation of these calibrated gray/dark pair-solitonic bi-trains
are generic collision phenomena of moderately/fully fragmented BEC,
which is strongly depending upon the equilibrium values of the red/blue-detuned
HGdT potential depth, respectively.

The dynamics of one gray/dark soliton in a BEC cloud is well described
by $\ddot{z}=-(1/2)\partial V_{\text{ext}}/\partial z$, where $V_{\text{ext}}$
is the dimensionless confining potential and $z$ denotes the position
of the gray/dark soliton. In the case of harmonic confinement with
a potential $V_{\text{ext}}=z^{2}/2$ the solution of this evolution
equation leads to an oscillation of the soliton described by $z(t)=\text{R}_{\text{TF1}}\sin\left(t/\sqrt{2}\right)$.
Thus the frequency of the oscillating soliton and the frequency of
the dipole oscillation of the Bose-Einstein condensate in the trap
differ by the factor $\sqrt{2}$ \cite{PhysRevLett.87.010401}. In
our system, pair-solitons bi-trains generally oscillate with the average
frequency $\Omega=2\pi\times4.80~{\rm Hz}$ irrespective of the sign
and the size of $U$ as shown in Fig.~\ref{Fig5}. With this, we get
the ratio $\Omega/\omega_{{\rm z}}\approx0.705$, which is quite close
to the dimensionless soliton frequency $1/\sqrt{2}\approx0.707$ in
a harmonic trap as predicted in Ref. \cite{PhysRevLett.87.010401}.
Note that previously the generation of solitons was studied theoretically
by investigating the collision of two condensates \cite{Scott98}
and experimentally for different quasi one-dimensional trap geometries
\cite{PhysRevLett.101.130401,Shomroni09}. Although in the latter
experiments only one potential maximum occurs instead of two as in
our work, so there single solitons and here pairs of solitons are
observed, the basic physics is the same. 

\begin{figure}
\includegraphics[scale=0.4]{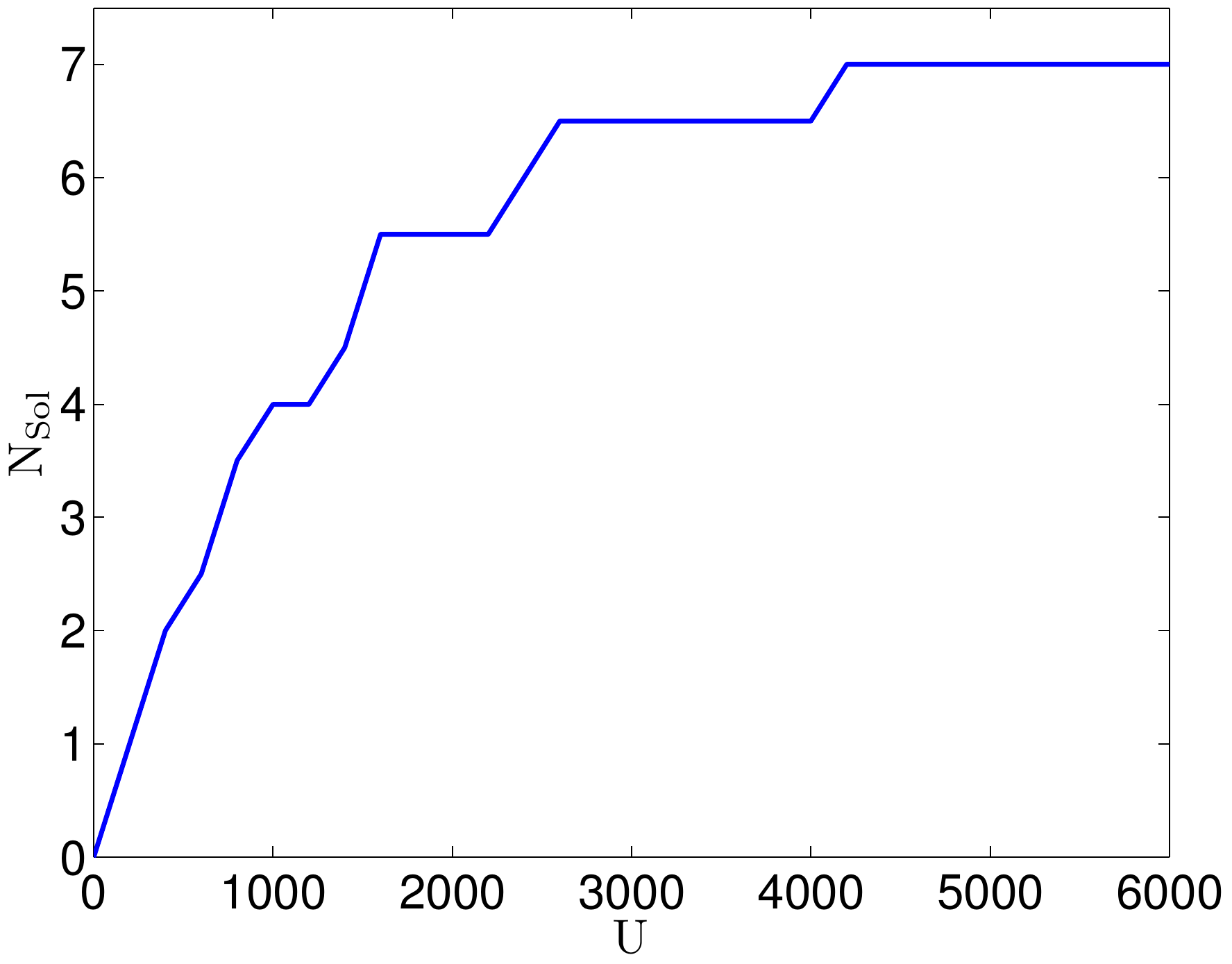}

\protect\protect\protect\protect\caption{(Color online) Number of pair-solitons $\text{N}_{\text{Sol}}$ in
each solitonic train versus HGdT potential depth $U$ in dimensionless
units. \label{Fig6} }
\end{figure}

We also observe an intriguing substructure of each soliton, which
we call pair-soliton. Normally, we find that there are solitons which
always move in pairs, and the mean distance between each other is
less than the neighboring solitons as shown in Fig.~\ref{Fig5}(b-d)
and Fig.~\ref{Fig5}(f-h). Numerically, we have observed that the
averaged distance between pair-solitons is less for dark solitons
as compared to the gray solitons as shown in Fig.~\ref{Fig5}(f-h).
We also observe that, in general, a minimal time of about $4.6\,$ms
is required to generate shock-waves/pair-solitons bi-trains as shown
in Fig.~\ref{Fig5}. The number of shock-waves is not effected by
the red/blue-detuned HGdT potential depth, but the number of interference
fringes increases. On the other hand, we observe that the number of
gray/dark pair-solitons depends on the depth of the red/blue-detuned
HGdT potential as shown in Fig.~\ref{Fig6}, the highest number of
pair-solitons in every train is 7. For the blue-detuned HGdT depth
$U<U_{\text{c}}$, the number of pair-solitons grows
linearly in the condensate and after the critical value $U_{\text{c}}$,
the number of pair-solitons remains approximately constant.\textbf{ }

Note that in case of the collision of two condensates in Ref. \cite{Scott98},
it turned out that the number of observable solitons depends sensitively
on the initial phase difference of both condensates. Thus, if the
two condensates have an initial phase difference of 0($\pi$), the
number of solitons is even(odd). In our case, we have a single BEC
fragmenting into three parts, which have the same phase, therefore
we observe an even number of pair-solitons in the condensate in agreement
with Ref. \cite{Scott98}. Indeed, Fig.~\ref{Fig6} shows the number
of pair-solitons in each solitonic train, so the total number of pair-solitons
in the whole condensate is twice as large. But, in Fig.~\ref{Fig6}
it turns out that the number of pair-solitons depends crucially on
the depth of the red/blue-detuned HGdT potential as shown in Fig.~\ref{Fig6}.

\section{Summary and Conclusion }
\label{Sec6}
We have developed a simple quasi 1D model both analytically and numerically
to calculate the statics and dynamics of the red/blue-detuned HGdT
imprint upon the $^{87}\text{Rb}$ condensate. First of all, we showed
a quantitative comparison between the Thomas-Fermi approximation and
numerical solutions for the underlying 1D Gross-Pitaevskii equation
for the equilibrium properties of the proposed system. Later we discussed
that the HGdT potential imprint upon the condensate wave function
strongly depends upon whether the effective HGdT is red- or blue-detuned.
With this we found out that the HGdT imprint generates two bumps/dips
at the center of the BEC density of the red/blue-detuned HGdT. Later
we discussed that the red-detuned HGdT induced two bumps did not decay
when we switched off the harmonic trap but the blue-detuned HGdT induced
two dips decay.\textbf{ }During the time of flight, we saw the emergence
of matter-wave stripes at the top of the condensate, which arise as
the BEC decomposes into a fraction at rest in the center and two moving
condensates at the borders. We have used the quasi one-dimensional
time-dependent GPE to analyze the creation of gray/dark pair-solitons
bi-trains within the moderately/fully fragmented BEC, which is strongly
depending upon the HGdT potential depth. The Hermite-Gaussian dimple
trap geometry maybe more applicable to soliton interferometry rather
than the Gaussian barrier adopted in Refs. \cite{Nguyen14,Javed15},
because one can shape solitons. Additionally, we also showed that
the number of pair-solitons in the system is depending on the initial
HGdT potential depth $U$. During the generation of pair-solitons
it was astonishing to find that the special shape of the newly generated
solitons in the harmonically trapped BEC is sculptured by the external
potential and the generation of gray/dark pair-solitons bi-trains
is a generic phenomenon of collisions of moderately/fully fragmented
BEC. With this we conclude that it maybe possible in the future to
frame complex shapes of solitons in the harmonically trapped BEC by
imposing a unique geometrical configuration for the external potential.

The ability of sculpting a quasi one-dimensional harmonic trapped
Bose-Einstein condensate by a HGdT has many exciting prospects. For
instance, it can be used to generate a truly continuous atom laser,
which has many applications in atom interferometry \cite{RevModPhys.81.1051,Buning10}.
To construct such an atom laser one needs a device that continuously
converts a source of condensed atoms into a laser-like beam. In Sec.
IV, we saw in the time-of-flight picture for the case $U>U_{\text{c}}$
that a BEC reservoir occurs at the center of the trap. By suitably
tuning the HGdT depth a fraction of this fragmented condensate could
be coupled out, serving as a source for an atomic beam.

\section{Acknowledgment}

We thank Thomas Bush for insightful comments. Furthermore, we gratefully
acknowledge financial support from the German Academic Exchange Service
(DAAD). This work was also supported in part by the German-Brazilian
DAAD-CAPES program under the project name ``Dynamics of Bose-Einstein
Condensates Induced by Modulation of System Parameters\textquotedblright{}
and by the German Research Foundation (DFG) via the Collaborative
Research Center SFB/TR49 ``Condensed Matter Systems with Variable
Many-Body Interactions\textquotedblright .

\bibliographystyle{apsrev4-1}
\bibliography{v1}

\end{document}